\documentclass[preprint,letterpaper,amsmath,superscriptaddress,floatfix]{revtex4}
\usepackage{amssymb,amsfonts,amsmath}
\usepackage{graphicx}
\usepackage{tabularx}
\begin{document}  %% Titlepage
%%%%%%%%%%%%%%%%%%%%%%%%%%%%%%%%%%%%%%%%%%%%%%%%%%%%%%%%%%%%%%%%%%%%%%%%%%%%%%%

\newcommand{\hm}{H$_3^-$}
\newcommand{\hp}{H$_3^+$}
\newcommand{\hdot}{$\text{H}^-\cdot\cdot\cdot \text{H}_2$}
\newcommand{\five}{RbH$_5$}

\title{Rubidium Polyhydrides Under Pressure: Emergence of the Linear H$_3^-$ Anion}
\author{James Hooper}
\author{Eva Zurek}\email{ezurek@buffalo.edu}
\affiliation{Department of Chemistry, State University of New York at Buffalo, Buffalo, NY 14260-3000, USA}
\begin{abstract}
The structures of compressed rubidium polyhydrides, RbH$_n$ with $n>1$, and their evolution under pressure are studied using density functional theory calculations. These phases, which start to stabilize at only  $P=2$~GPa, consist of Rb$^+$ cations and one or more of the following species: H$^-$ anions, H$_2$ molecules, and H$_3^-$ units. The latter motif, the simplest example of a three--center four--electron bond, is found in the most stable structures, RbH$_5$ and RbH$_3$.  Pressure induces the symmetrization of H$_3^-$. The thermodynamically most stable polyhydrides metallize above 200~GPa. At the highest pressures studied, our evolutionary searches find an RbH$_6$ phase which contains polymeric (H$_3^-$)$_\infty$ chains that show signs of one--dimensional liquid--like behavior.
\end{abstract}
%\pacs{71.20Dg, 62.50.-p, 63.20.dk}% PACS, the Physics and Astronomy Classification Scheme.
%Electron density of states and band structure of crystalline solids
%http://www.aip.org/pacs/pacs2010/individuals/pacs2010_regular_edition/reg70.htm#71
%71.20.Ps	Other inorganic compounds
%%71.18.+y	Fermi surface: calculations and measurements; effective mass, g factor
\keywords{high pressure, alkali hydrides, metallization, first principles}
\maketitle

\section{Introduction} \label{sec:intro}

\hp\ has been known to mass spectroscopists since Thomson reported it in 1912 \cite{Thomson:1912a}, but
it took 90 more years to reliably observe \hm\ \cite{Golser:2005a, Gnaser:2006a, Wang:2003a}. Already in 1935 Coulson predicted the triangular configuration of the cation, and the linear arrangement of its anionic cousin \cite{Coulson:1935a}. Theoretical studies confirmed these geometries \cite{Schwartz:1967a} but also concluded that H$_3^-$ was unstable with respect to decomposition into H$_2$ and H$^-$ \cite{Stevenson:1937a}. In the 1990's \textit{ab--initio} computations became reliable enough to show that H$_3^-$ is a minimum on the potential energy surface (PES) \cite{Starck:1993a, Robicheaux:1999a}, which has been subsequently explored \cite{Ayouz:2010a, Belyaev:2006a}. We came across the H$_3^-$ anion serendipidously during our computational quest to predict hydrogen--rich structures that are metallic and potentially superconducting at experimentally accessible pressures \cite{Ashcroft:1968a, Ashcroft:2004a}.  

Hydrogen turns black at 320~GPa \cite{Loubeyre:2002}, but it does not become metallic even at pressures approaching those found in the Earth's core \cite{Narayana:1998a}. Once metallized, H$_2$ is predicted to become superconducting at temperatures as high as 240~K \cite{Cudazzo:2008a}. On the other hand, band gap closure and concommitant superconductivity in various metal hydrides such as  GeH$_4$ \cite{Gao:2008a}, SnH$_4$  \cite{Tse:2007a}, AlH$_3$ \cite{Pickard:2007b, Goncharenko:2008a}, and GaH$_3$ \cite{Gao:2011a} has been computed for pressures attainable in diamond anvil cells. Various attempts to dope hydrogen by an impurity cabable of promoting metallization have been carried out \cite{Klug:2011a}. Compression not only affects the electronic structure and properties of extended systems, but it also modifies their \emph{chemistry}: it may entice elements which would not normally combine to do so \cite{Feng:2008a}, or to mix in non--traditional proportions. An example of the latter is the van der Waals compound Xe(H$_2$)$_7$  \cite{Somayazulu:2010a}, or the SiH$_4$/H$_2$ phases with ratios of 1:1, 1:2, and 1:5  \cite{Strobel, Wang:2009a} which have been observed in recent experiments. 

%and a superconducting SiH$_4$ + H$_2$ system was found at 96~GPa \cite{Wang:2009a}; efforts to metallize Xenon + H$_2$ resulted in the discovery of a unique hydrogen--rich lattice at 4.8 GPa \cite{Somayazulu:2010a}, but the structure remained semi-conducting up to 250~GPa. A number of materials of been investigated computationally, predicting, for example, superconducting PtH crystals under pressure \cite{Kim:2011a} and stable and metallic Lithium and Sodium polyhydrides \cite{Zurek:2009c, Zurek:2011d}.  

The exciting findings in high--pressure research have inspired us to theoretically explore the potential energy landscape of compressed alkali metal polyhydrides, MH$_n$ with $n>1$. These stoichiometeries are unstable with respect to decomposition into MH and H$_2$ at atmospheric conditions, but calculations show that the enthalpies of formation of the most stable lithium \cite{Zurek:2009c} and sodium polyhydrides \cite{Zurek:2011d} become negative at about 100 and 25~GPa.  The pressures necessary to stabilize these systems were found to correlate with the ionization potentials (IP) of the metals, and the preferred stoichiometries (LiH$_6$  and NaH$_9$)  with the size of the alkali metal cation. 

Rb has an even lower IP than Na, and Rb$^+$ a larger radius than Na$^+$. In order to determine how this affects the rubidium polyhydrides, we have performed first--principles calculations coupled with the evolutionary algorithm (EA) XtalOpt \cite{Zurek:2011a, Zurek:2011f} to search for the most stable RbH$_n$ up to 250~GPa. The results of our studies on RbH$_n$ ($n>1$) illustrate that under pressure: (i) compounds with unusual stoichiometries may be stabilized; (ii) there is an increased tendancy to form multicenter bonds; and (iii) orbitals which chemists do not usually think of (ie.\ Rb $4s$ and $4p$) become important. 

Below we explore the various phases of the rubidium polyhydrides at pressures ranging from 2~GPa --- the onset of stabilization ---  up to the eventual insulator--to--metal transition (MIT) which occurs just over 200~GPa. RbH$_5$, the most stable structure throughout a large part of the pressure range considered, contains H$_2$ molecules, Rb$^+$ cations and H$_3^-$ anions which become symmetric above 30~GPa. Other fascinating phases, such as RbH$_9$ and RbH$_3$ also have domains of thermodynamic stability. At the greatest compressions considered, the hydrogen sublattice itself shows signs of polymerization.

%Our calculations indicate that RbH$_n$ does indeed stabilize at lower pressures, near 2~GPa. At 10~GPa, its preferred stoichiometry of RbH$_9$ consists of Rb$^+$, H$^-$, and H$_2$, similar to NaH$_9$ above NaH$_n$'s domain of existence. At higher pressures, however, RbH$_n$ behaves differently than its lighter alkali counterparts. Its preferred stoichiometry of RbH$_5$ at 100~GPa has less hydrogen than either Li or Na and, in addition, adopts a crystal structure with distinct $H_{3}^{-}$ subunits instead of hydridic hydrogen atoms. 

%. Upon calculating the stability of RbH $+$ $0.5(n-1)$H$_2 \rightarrow$ RbH$_{n}$ with respect to a number of $n$ values up to 250~GPa, we find further stabilization of the H$_3^-$ molecule in a RbH$_3$ stoichiometry. We find the MIT transition occurs 

\section{Results and Discussion} \label{sec:discussion}

\begin{figure} [t!]
\begin{center}
\includegraphics[width=\columnwidth]{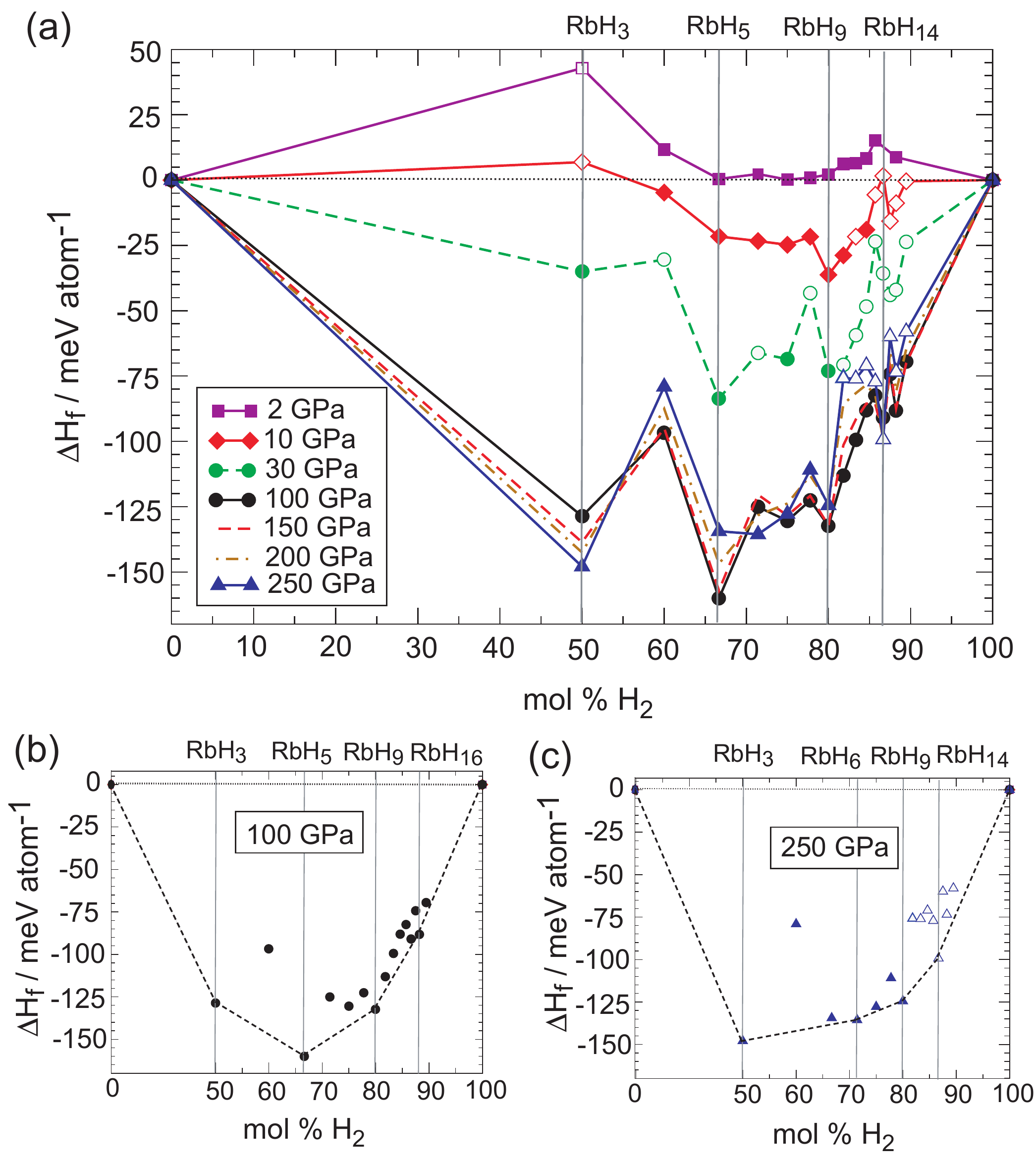}
\end{center}
\caption{(a) The enthalpy of formation, $\Delta H_{F}$, for the reaction $\text{RbH} + \frac{1}{2}(\text{H}_2)_{n-1} \rightarrow \text{RbH}_n$ versus the H$_2$ composition at varying pressures, where the enthalpy of RbH and H$_2$ are computed for the most stable structures from Refs.\ \cite{Ahuja:1999a} and \cite{Pickard:2007a}. The corresponding convex hulls at (b) 100~GPa and (c) 250~GPa are shown below to highlight thermodynamically stable RbH$_n$ structures. The horizontal axes in (b) and (c) are not to scale. Filled symbols represent points at which EA structure searches were explicitly run; open symbols represent structures which were re-optimized at said pressures from an EA search performed at another pressure. 
\label{fig:RbHN_tieplot}}
\end{figure}

The calculated enthalpies of formation, $\Delta H_F$, of the most favorable RbH$_n$ structures found in our evolutionary runs are provided in Fig.\ \ref{fig:RbHN_tieplot}(a). RbH$_2$ (33.3 mol\% H$_2$) is not shown since it was the only phase found to be unstable with respect to decomposition into RbH and H$_2$ at every pressure considered.  At only $\sim$2~GPa RbH$_n$ with $n$=5--9 become preferred. This industrially attainable pressure is an order of magnitude lower than that necessary to stabilize the sodium polyhydrides, highlighting again the import of the alkali metal's IP for the formation of these phases.

If a tie--line is drawn connecting $\Delta H_F$(A) with $\Delta H_F$(B) and $\Delta H_F$(C)  falls beneath it, then A and B will react to form C. If another phase D falls above the tie--line connecting A and B, it is expected to decompose into these two structures, provided the kinetic barrier is not too high. The tie--lines may be used to construct a convex hull --- a point lying on it corresponds to a thermodynamically stable phase at the given pressure \cite{Feng:2008a}. The dashed line in Fig.\ \ref{fig:RbHN_tieplot}(b) traces out the estimated convex hull of RbH$_n$ at 100~GPa, where RbH$_5$ has the most negative enthalpy of formation. Other thermodynamically stable phases are RbH$_3$, RbH$_9$, and RbH$_{16}$. Above this pressure the $\Delta H_F$ of all of the phases --- with the exception of RbH$_3$ --- increases. The destabilization of structures at greater compression was not previously observed in studies of the lighter alkali polyhydrides up to 300~GPa \cite{Zurek:2009c, Zurek:2011d}. At 250~GPa RbH$_3$ is the most stable structure, with RbH$_6$, RbH$_9$ and RbH$_{14}$ falling on the convex hull. Comparison of these results with previous findings for LiH$_n$ and NaH$_n$ suggests that there is no simple relationship between the preferred stoichiometries of the alkali metal polyhydrides and the radius of the cation.  

Within the pressure range considered, the most stable phases contain between 3 and 9 hydrogen atoms per rubidium. They may all be reasonably divided into four groups which are based on qualitative assessments of the atoms' spatial arrangements:  1) Rb$^+$ cations and (H$_{2}$)$^{-n/2}$ molecules with bonds longer than 0.75~\AA{}; 2) Rb$^+$, H$_2$ molecules, and hydridic H$^-$ anions; 3) Rb$^+$, H$_2$ and H$_3^-$ molecules; and 4) Rb$^+$ and H$_3^-$. The first two structural makeups have already been observed in LiH$_n$ and NaH$_n$. The latter two groups have not yet been noted, but are found in the most stable rubidium polyhydrides, RbH$_5$ and RbH$_3$. RbH$_6$ consists of arrangements of H$_3^-$ units brought together to form what are essentially chains. Herein, we examine the geometries of these phases, and their electronic structures --- especially the imminent metallization under compression.

\subsection{Pressure--Induced Stabilization of H$_3^-$} \label{subsec:lowpressure}

\begin{figure} [b!]
\begin{center}
\includegraphics[width=0.8\columnwidth]{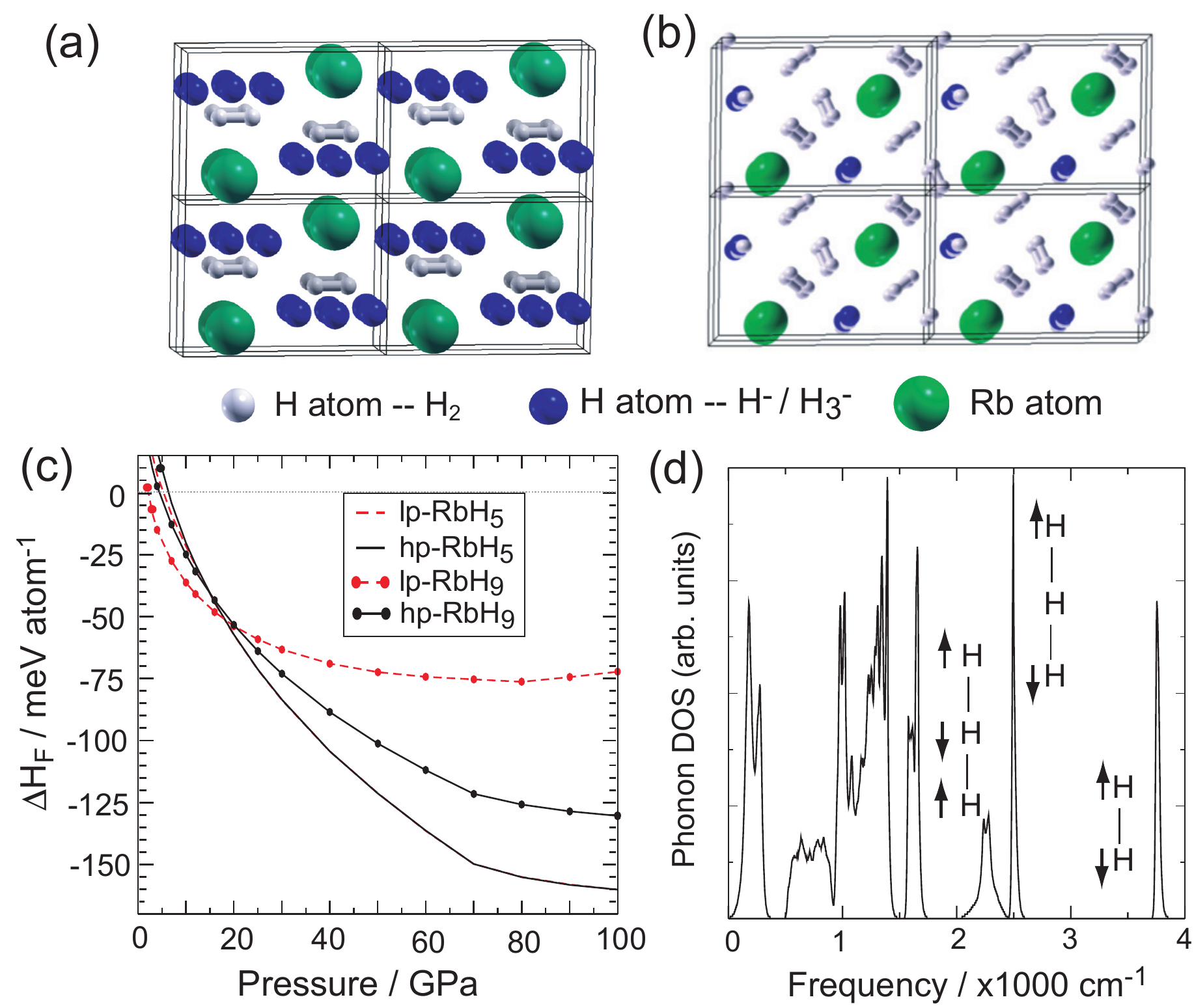}
\end{center}
\caption{$2\times 2 \times 2$ supercells of the (a) hp--RbH$_5$ phase at 100~GPa, and the (b) lp--RbH$_9$ phase at 10~GPa. The hydrogen atoms with H$^-$ or H$_3^-$ character are shaded to enhance visibility. (c) $\Delta H_F = H(\text{RbH}_n)-(\frac{n-1}{2}) H(\text{H}_2) -H(\text{RbH})$ for these structures. (d) Phonon calculations of \five\ at 30 and 100~GPa did not reveal any imaginary frequencies, indicating that these are local minima. The phonon DOS for hp--RbH$_5$ at 100~GPa is shown.} \label{fig:hooper-fig1}
\end{figure}

Below 100~GPa the most stable stoichiometries are RbH$_9$ and RbH$_5$. The enthalpy of formation of a  $Pm$--symmetry RbH$_9$ phase  becomes negative near $\sim$2~GPa and RbH$_9$ remains the preferred combination up to $\sim$16~GPa, at which point it is overtaken by RbH$_5$ as shown in Fig.\ \ref{fig:hooper-fig1}(c). The lowest enthalpy structures differ at 10 and 100~GPa, so below we refer to them as the ``low pressure'' (lp) and ``high pressure'' (hp) structures.  

The linear motifs which characterize the \hm\ units within the $Cmcm$--symmetry hp--RbH$_5$ phase are highlighted in Fig.\ \ref{fig:hooper-fig1}(a) \footnote{At 100~GPa: $a$=2.97~\AA{}, $b$=7.39~\AA{}, $c$=4.88~\AA{} with the atoms at the following Wyckoff sites: H $4f$ (0.500   0.781   0.171), H $4f$ (0.500   0.360   0.064), H $2c$ (0.500   0.633   0.750), and Rb $2c$ (0.500   0.923   0.750)}. At 100~GPa, the intramolecular H--H distances in the H$_3^-$ and H$_2$ units measure 0.91~\AA{} and 0.77~\AA{}. The latter is slightly elongated when compared with the 0.73~\AA{} in pure $P6_3/m$--H$_2$ \cite{Pickard:2007a} at this pressure. The shortest intermolecular H--H distances in hp--\five\ are roughly equivalent to those in pure H$_2$ (1.56~\AA{} vs.\ 1.50~\AA{}) implying that RbH$_5$ can reasonably be thought of as containing an arrangement of H$_2$ and H$_{3}^{-}$ molecules. 

From visualization of the phonon modes of hp--RbH$_5$ lying above 2000~cm$^{-1}$, they are assigned to H$_3^-$ molecular vibrations between 2000--2700~cm$^{-1}$ and H$_2$ near 4000~cm$^{-1}$. These modes are shown schematically in the phonon DOS in Fig.\ \ref{fig:hooper-fig1}(d), and could serve as a fingerprint to characterize RbH$_5$, particularly the symmetric H$_3^-$ stretch near 2500~cm$^{-1}$.  

The symmetric H$_{3}^{-}$ units in hp--RbH$_5$ persist as the pressure is decreased until 15~GPa, wherein the system becomes thermodynamically unstable with respect to lp--RbH$_5$ with asymmetric \hdot\ units. Actually, at around this same pressure both RbH$_5$ phases are predicted to decompose into RbH and RbH$_9$.  The lp--RbH$_9 $ shown in Fig.\ \ref{fig:hooper-fig1}(b) at 10~GPa also adopts asymmetric \hdot\ arrangements but, unlike lp--RbH$_5$, each H$^-$ atom has several weak (linear) \hdot\ interactions. At 10~GPa, the four closest \hdot\ contacts in lp--RbH$_9$ measure $\sim$1.85~\AA{} but there is only one such contact in lp--RbH$_5$ measuring 1.27~\AA{}. 
%At only $\sim$2~GPa all of the RbH$_n$ phases become unstable with respect to decomposition into RbH and H$_2$,  highlighting again the import of the alkali metal's IP on the pressure at which the polyhydrides become viable. 

The \hdot\ arrangements adopted in these structures conform to the expectation of an increase in multicenter bonding in compressed solids \cite{Grochala:2007a}. Such a precedence has already been set in pure H$_2$ wherein calculations predict that at a few TPa triangluar \hp\ motifs may form along with a background of negative charge \cite{Mcmahon:2011a}. The electronic structure of \hp\ and \hm\ can be understood in terms of three--center bonding schemes; for two electrons (3c-2e) electron poor, and four electrons (3c-4e) electron rich. 

A classic example of a 3c-4e bond is the triiodide anion. In fact Rb forms one of the salts of I$_3^-$ \cite{Tebbe:1986a} which, if you consider $\text{Rb}^+\text{H}_3^-$ as an alkali trihalide, also agrees with the observation that pressure can favor configurations adopted by heavier elements in the same group \cite{Grochala:2007a}.   

\begin{figure} [h!]
\begin{center}
\includegraphics[width=0.8\columnwidth]{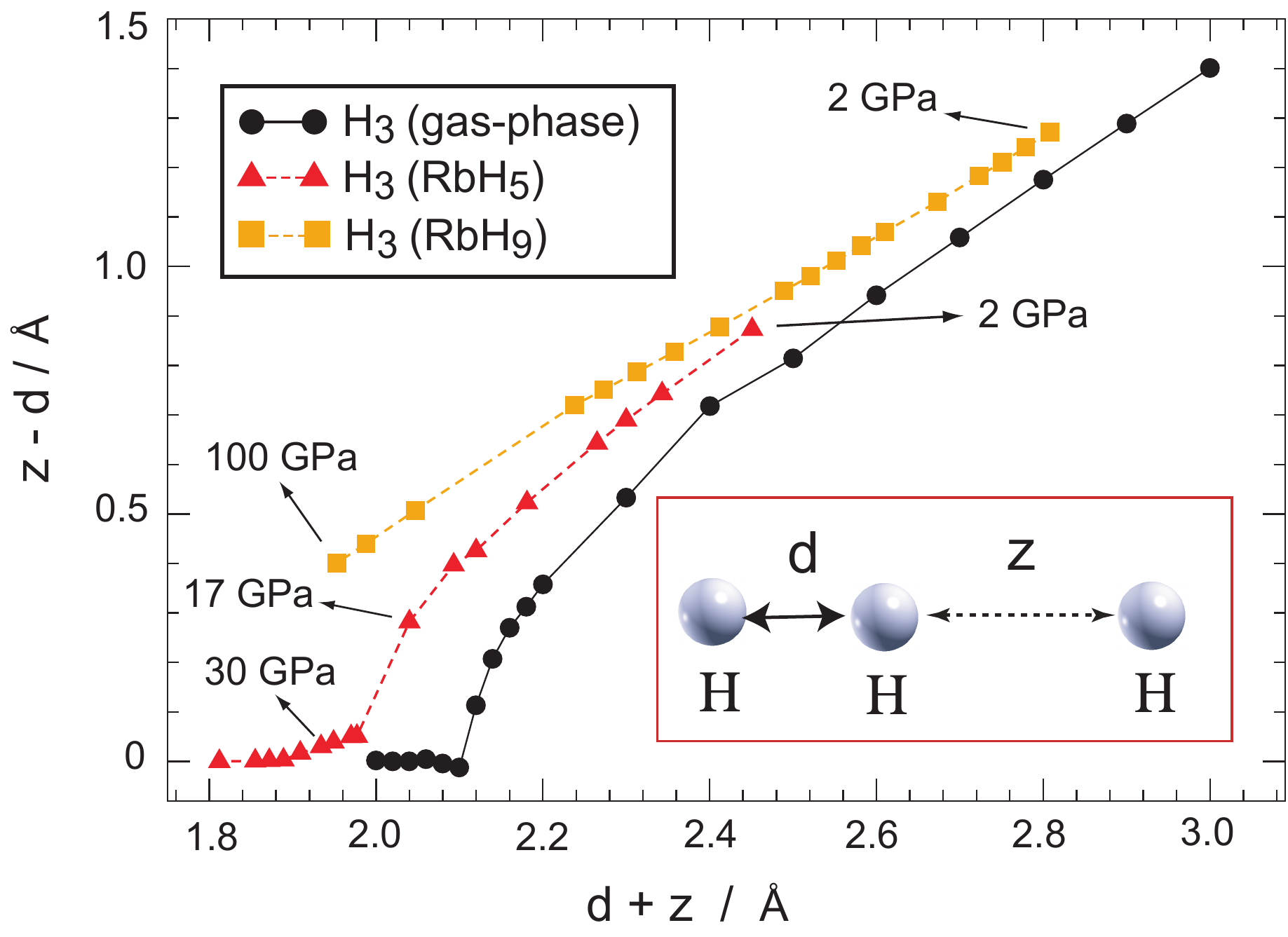}
\end{center}
\caption{The difference between the $z$ and $d$ distances in an \hdot\ fragment (see inset) is plotted vs.\ their sum for an isolated H$_3^-$ molecule in the gas phase. During the optimization, the total length of the molecule was constrained and the central atom was allowed to relax freely. These values are also illustrated for an \hdot\ unit in lp--RbH$_9$ and lp--RbH$_5$. These phases were reoptimized in order to determine the effect of pressure on ($d+z$) and ($d-z$).\label{fig:hooper-fig3}}
\end{figure}

The shallow energy minima in the H$_3^-$ PES yields H--H distances ($d$ and $z$ depicted in Fig.\ \ref{fig:hooper-fig3}) measuring 0.75 and 2.84~\AA{} \cite{Ayouz:2010a}. The lowest energy barrier connecting the two minima is a symmetric H$_3^-$ molecule with H--H distances of $\sim$1.06~\AA{} \cite{Ayouz:2010a}. Such symmetric H$_3^-$ motifs have also been discussed in the literature as transition states in hydrogen exchange processes of M(H)(H$_2$) metal complexes \cite{Maseras:2000a}.

The observation of symmetric H$_3^-$ units at 100~GPa and their asymmetrization as the pressure is lowered ($\sim$10~GPa) suggests that the equalization of $d$ and $z$ is pressure--driven. To test this further a computational experiment was carried out wherein the total length ($d+z$) of an \hdot\ unit was constrained to a fixed value in order to mimic an applied external pressure, and the molecule was subsequently optimized \footnote{The molecular calculations on H$_3^-$ were performed using the ADF (www.scm.com) package (BP functional with DZP basis set )}. In Fig.\ \ref{fig:hooper-fig3} we give the sum of the H--H distances on the horizontal axis, and their difference along the vertical axis. When no constraints were imposed, the equilibrium configuration was such that $d$=0.81~\AA{} and $z$=1.97~\AA{}. As the total length of the molecule was shortened, both H--H distances gradually approached each other until equalization was observed when ($d+z$) fell below 2.1~\AA{}.   

For comparison we also provide data extracted from lp--RbH$_5$ when it was reoptimized at intermittent pressures between 2-100~GPa. Between 17-20~GPa, equalization is observed when the total length of the H$_3^-$ motifs nears 2.0~\AA{}. The onset of equalization occurs at very similar ($z+d$) values in both the molecule and in the solid. 
 
The lp--RbH$_9$ structure also has asymmetric \hdot\ arrangements. However, equalization is not achieved at 100~GPa in spite of ($d+z$) falling below 2.0~\AA{}. We attribute this behavior to RbH$_9$'s structural peculiarities: whereas in RbH$_5$ each H$^-$ has only one H$_2$ contact, in RbH$_9$ there are four. As a result, the H$_2$ unit is substantially shorter in RbH$_9$ than in RbH$_5$ (it measures $\sim$0.85~\AA{} in lp--RbH$_5$, but never surpasses 0.78~\AA{} in RbH$_9$ even at 100~GPa). The total length of the fragment is substantially shorter at 2~GPa for RbH$_5$ than it is for RbH$_9$ suggesting that in the former phase the hydrogen sublattice is more compressed.

\begin{figure} [h!]
\begin{center}
\includegraphics[width=0.8\columnwidth]{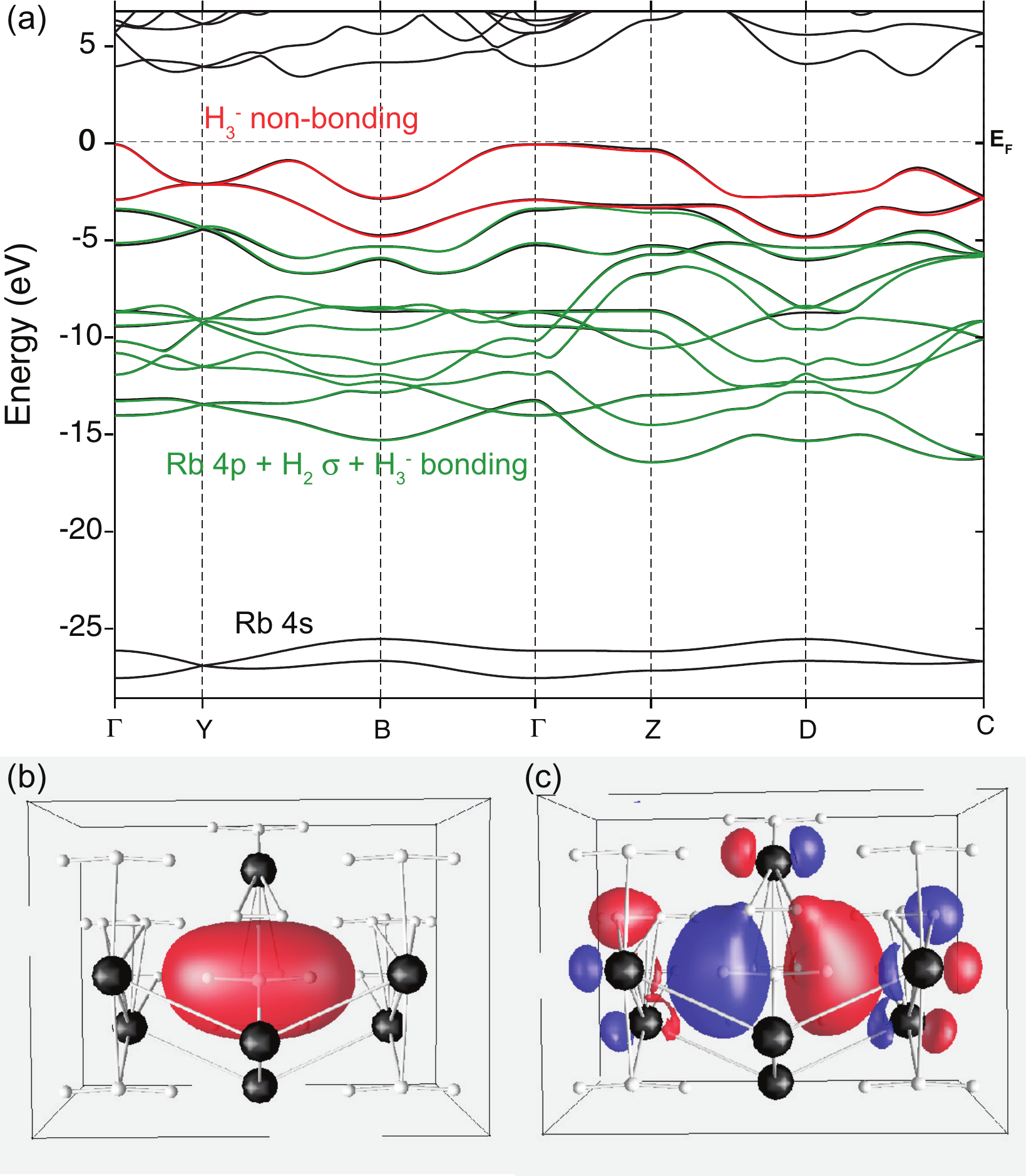}
\end{center}
\caption{(a) The bands of $Cmcm$ hp--RbH$_5$ at 100~GPa calculated with a full Muffin Tin Orbitals of order $N$ ($N$MTO)  basis %(Rb $4s$, $4p$, $5d$ and H $1s$, $2p$) 
are given in black. The green bands were computed with a minimal basis describing the Rb $4p$, H$_2$--$\sigma$, and \hm\ HOMO-1 bands, and the red bands the \hm\ HOMO. Along these high--symmetry lines the red and green bands do not cross. We illustrate the \hm\ centered WFs obtained from the basis sets which yielded the (b) green, and (c) red bands.\label{fig:hooper-fig4}}
\end{figure}

If sufficiently squeezed the core electrons of the alkali metals start to overlap, the bands broaden and the valence electrons become localized in the interstitial regions; Na, for example, becomes an insulator at $\sim$200~GPa \cite{Ma:2009a}. For K, Rb and Cs, pressure is also known to induce an electronic $s\rightarrow d$ transition, and by $\sim$50~GPa Rb is in essence a $d^1$ metal \cite{McMahan:1984a}. Compress further, and the $5d$ band hybridizes with the broadened $4p$ semicore states. In Ref.\ \cite{McMahan:1984a} it was shown that both the core repulsion (a result of the overlap), and the hybridization are important factors in determining the most stable elemental Rb structures.  
 
As illustrated in the Supporting Information (SI), the bands near the Fermi level in hp--RbH$_5$ at 100~GPa display some Rb $5d$ character. This is a minor contribution however, since the valence electron has been ionized into the hydrogen sublattice. Much more important for the electronic structure at this pressure is the overlap of the $4p$ semicore orbitals. In fact, at 100~GPa even the Rb $4s$ orbitals interact with each other, as evidenced by their 2.2~eV bandwidth. The substantial core overlap affects the electronic structure of hp--RbH$_5$: a calculation of the DOS of the (H$_5$)$^{-}$ sublattice shows this hypothetical system to be a good metal (see the SI). 

Because of their overlap and hydridization with the Rb $4p$ bands, it is not possible to construct a Wannier function (WF) which describes only the H$_2$--$\sigma$ bands. One can, however, construct a minimal basis set \cite{Zurek:2005a, Zurek:2010a} that exactly reproduces the ten bands whose energies fall in between -18 to -5~eV in Fig.\ \ref{fig:hooper-fig4}(a). The WFs obtained from this basis are the Rb $p$ non--bonding and H$_2$--$\sigma$ bonding  (both not shown), as well as the delocalized H$_3^-$ $\sigma$--bonding orbital illustrated in Fig.\ \ref{fig:hooper-fig4}(b). One may also obtain a WF which yields just the two red valence bands. This WF, provided in Fig.\ \ref{fig:hooper-fig4}(c), is primarily H$_3^-$ non--bonding, but it also contains some Rb $p$ and H$_2$--$\sigma$ character. The filling of the two orbitals illustrated in Fig.\ \ref{fig:hooper-fig4}(b, c) is a hallmark of a 3c-4e bond as would be found in the symmetric \hm , confirming our analysis of the bonding in hp--RbH$_5$.

As interesting as RbH$_5$ is, unfortunately it cannot be superconducting at least up to 100~GPa where the PBE band gap is $\sim$1.7~eV. In fact, the symmetrization of H$_3^-$ is likely detrimental to the metallization. We calculate the HOMO--LUMO gap of an H$_3^-$ molecule with $z=d=$~0.9~\AA{} as being 6.2~eV --- over 40\% larger than the 4.4~eV gap in the asymmetric fully optimized species. 

\subsection{Metallization of RbH$_n$}

The $\sim$13~eV HOMO--LUMO gap in pure H$_2$ remains large as the molecules are brought together and bands form. The $\sigma_g$--bonding and $\sigma_u^*$--antibonding bands undergo pressure--induced broadening, and are predicted to eventually overlap. But in experiments this does not happen below 342~GPa \cite{Narayana:1998a}.
\begin{figure} [b!]
\begin{center}
\includegraphics[width=\columnwidth]{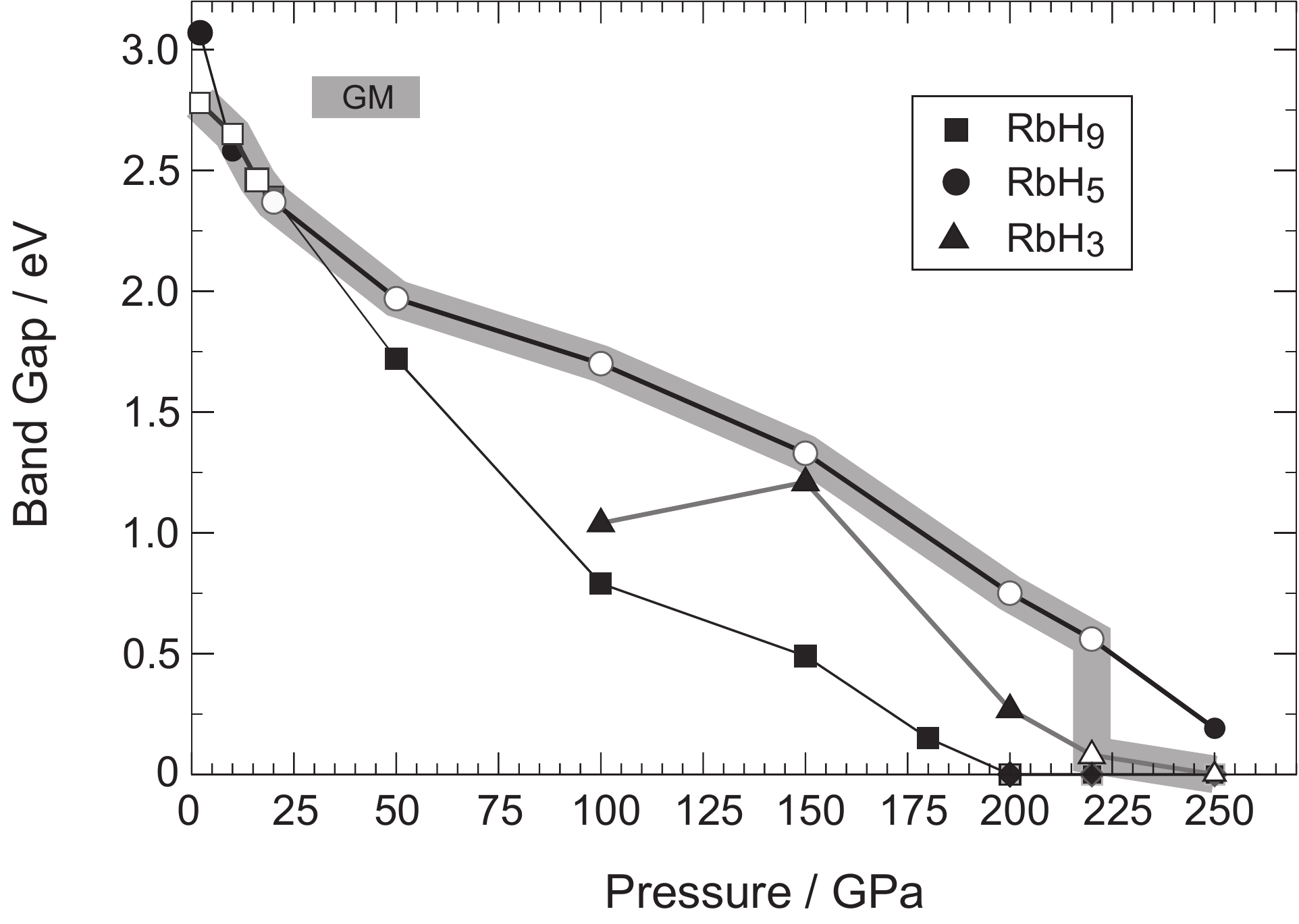}
\end{center}
\caption{The evolution of the PBE band gaps of the lowest--enthalpy RbH$_9$, RbH$_5$, and RbH$_3$ structures with increasing pressure. The most thermodynamically stable RbH$_n$ species, ie.\ the global minima (GM), are shown as unfilled symbols and are traced with a gray line. RbH$_6$ is also stable with respect to decomposition at 250~GPa, and is metallic throughout its domain of stability. 
\label{fig:bandgaps}}
\end{figure}%

One strategy to achieve metallization at experimentally achievable pressures is to perturb the electronic structure of H$_2$ by the addition of impurities \cite{Klug:2011a}. Doping with an electropositive element can lead to the partial filling of the H$_2$ $\sigma_u^*$ bands and metallicity already at 1 atm as in LiH$_6$ \cite{Zurek:2009c}. Unfortunately, LiH$_6$ remains unstable with respect to decomposition into H$_2$ and LiH below 110~GPa. Another possibility is the formation of an  H$^-$ donor--impurity band falling between the $\sigma_u$ and $\sigma_g^*$ bands as in LiH$_2$. But the overlap of the Li $1s$ core orbitals hinders metallization, so this system remains a stable semi--metal above 120~GPa. Even though the sodium polyhydrides become viable at $\sim$25~GPa, a metallic species which lies on the convex hull emerges only at 250~GPa. This NaH$_9$ phase consists of an Na$^+$/H$^-$ sublattice filled with H$_2$ molecular guests, and overlap of the Na $2p$ cores forestalls metallization.

The pressure--dependance of the PBE band gaps of RbH$_9$, RbH$_5$, and RbH$_3$ (species which lie on the convex hull at some point within 2-250~GPa) are shown in Fig.\ \ref{fig:bandgaps}. For all of these structures the calculated MIT transitions occur near 220~GPa, with the gap in RbH$_9$ closing at slightly lower, and in RbH$_5$ at slightly higher pressures. 

Two other structures, RbH$_7$ and RbH$_8$, were found to metallize by 100~GPa. However, their enthalpies of formation fell above the convex hull and their calculated phonon dispersions showed soft modes. At 150~GPa  P$4/nmm$--RbH$_7$ --- with Rb$^+$ cations, H$_2$ molecules, and \hdot\ moieties --- was found to be mechanically stable, but it's $\Delta H_F$ still lay above the convex hull.\\[2ex]

\begin{figure} [h!]
\begin{center}
\includegraphics[width=\columnwidth]{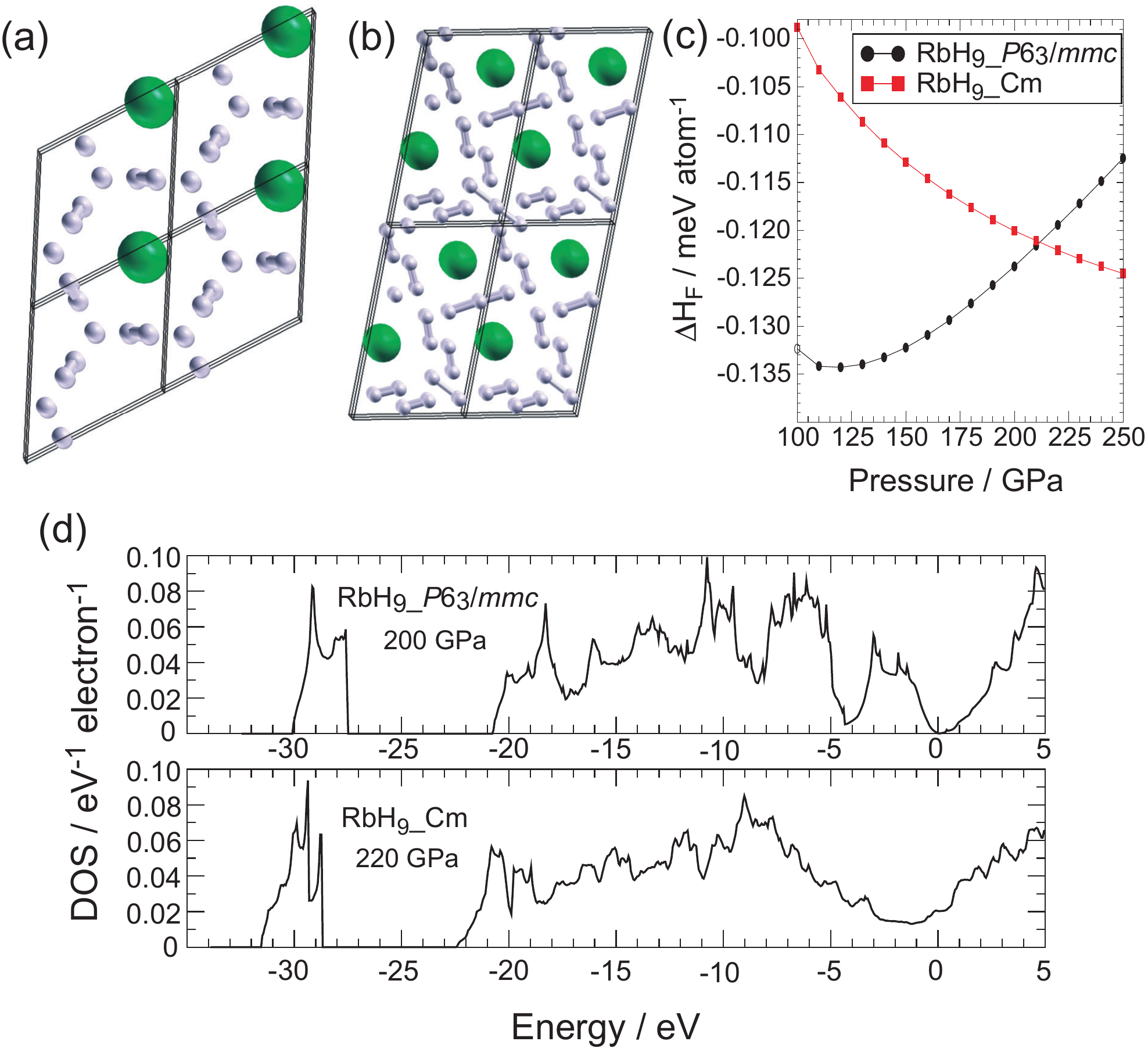}
\end{center}
\caption{Supercells of the lowest enthalpy RbH$_{9}$ phases recovered at  (a) 150~GPa ($P6_3/mmc$--symmetry), and (b)  250~GPa ($Cm$--symmetry); and (c) their enthalpies of formation at various pressures. (d) Total densities of states of (top) $P6_3/mmc$--RbH$_{9}$ at 200~GPa, and (bottom) $Cm$--RbH$_{9}$ at 220~GPa. The Fermi levels are set to zero.  
\label{fig:RbH9_Figure}}
\end{figure}

\noindent\textbf{Structures of RbH$_9$ and RbH$_5$ near the MIT} \\
The behavior of compressed RbH$_9$, which remains stable with respect to decomposition up to 250~GPa, is akin to NaH$_9$. Both systems can be thought to consist of an M$^+$/H$^-$ lattice within a sea of H$_2$ molecules, and the band gap between the occupied hydridic bands and the H$_2$ $\sigma_{u}^{*}$ bands eventually closes as the pressure is increased to $\sim$200~GPa. $P6_3/mmc$--RbH$_9$ at 150~GPa \footnote{At 150~GPa: $a$=3.65~\AA{}, $c$=5.42~\AA{} with the atoms at the following Wyckoff sites: H $12k$ (0.535   0.465   0.057), H $4f$ (0.333   0.667   0.317), H $2d$ (0.667   0.333   0.250), and Rb $2b$ (0.000   0.000   0.250)}, see Fig.\ \ref{fig:RbH9_Figure}(a), was confirmed to be mechanically stable based on the absense of imaginary phonon modes. 

The total electronic densities of states (DOS) of this phase at 200~GPa is provided at the top of Fig.\ \ref{fig:RbH9_Figure}(d). The projected densities of states (see the SI) confirm the broadening of the Rb $4s$ levels around -30~eV, the mixing of Rb $4p$ with the H$_2$ $\sigma_u$ bands, and the contribution of the hydridic ions to the valence band. In addition, the H$_2$--hydrogens furthest away from the H$^-$ atoms contribute more to the valence DOS than their nearest neighbors. This type of density distribution is consistent with the HOMO in a system containing an assymmetric 3c--4e bond (see the SI), confirming the interaction of the H$_2$ and H$^-$ entities via delocalized multicentered bonding.
 
At 220~GPa the enthalpy of $Cm$--RbH$_9$ recovered from an EA search at 250~GPa becomes lower than that of the hexagonal phase (see Fig.\ \ref{fig:RbH9_Figure}(c)). Examination of the extended structure shown in Fig.\ \ref{fig:RbH9_Figure}(b) may lead one to classify it as an arrangement of Rb$^+$ cations, H$_2$ molecules, and slightly asymmetric H$_3^-$ fragments. The second--nearest neighbor distance about each hydrogen atom in $Cm$--RbH$_9$ lies between 0.93-1.16~\AA{}, as compared with 1.27-1.38~\AA{} in the $P6_3/mmc$--symmetry structure at 220~GPa. The more tightly--packed $Cm$--phase has a slightly lower volume, by $\sim$2\% at 220~GPa, and is therefore preferred at higher compression. This phase is  metallic with a much higher density of states at the Fermi level than the hexagonal structure. Such a scenario is reminiscent of the sodium polyhydrides for which the enthalpy of a metallic $Cmc2_1$--NaH$_9$ structure became lower than that of insulating $Pm$--NaH$_9$ at 300~GPa. However, due to geometry convergence issues, we were unable to confirm the mechanical stability for $Cm$--RbH$_9$.  
\begin{figure} [h!]
\begin{center}
\includegraphics[width=\columnwidth]{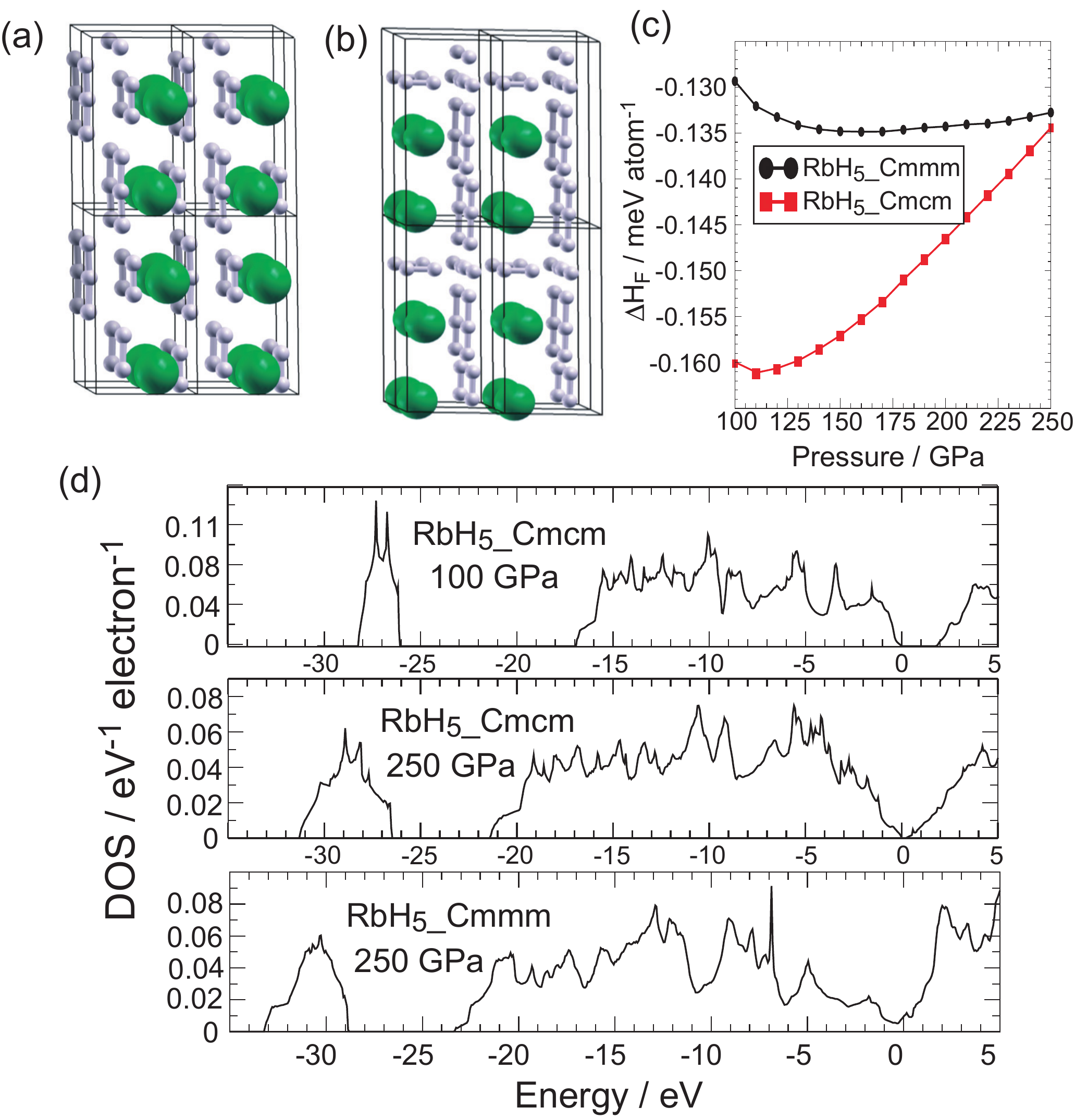}
\end{center}
\caption{Supercells of the lowest enthalpy RbH$_{5}$ phases recovered at  (a) 100~GPa ($Cmcm$--symmetry), and (b)  250~GPa ($Cmmm$--symmetry); and (c) their enthalpies of formation at various pressures. (d) Total densities of states of (top) $Cmcm$--RbH$_{5}$ at 100~GPa, (middle) $Cmcm$--RbH$_{5}$ at 250~GPa, and (bottom) $Cmmm$--RbH$_{5}$ at 250~GPa. The Fermi levels are set to zero.  
\label{fig:RbH5_Figure}}
\end{figure}

The $Cmcm$ hp--RbH$_5$ phase illustrated in Fig.\ \ref{fig:hooper-fig1}(a) and reproduced in Fig.\ \ref{fig:RbH5_Figure}(a) remains the most stable RbH$_5$ structure up to 250~GPa. In going from 100 to 250~GPa, its band gap decreases from to 1.7 to 0.2~eV, see Fig.\ \ref{fig:RbH5_Figure}(d). The convex hull, Fig.\ \ref{fig:RbHN_tieplot}(c), shows that RbH$_5$ becomes slightly unstable with respect to decomposition into RbH$_6$ and RbH$_3$ at 250~GPa.  The EA search at 250~GPa also recovered a metallic $Cmmm$--RbH$_5$ arrangement shown in Fig.\ \ref{fig:RbH5_Figure}(b). The two structures possess very similar interatomic distances, and are within dynamical enthalpies of each other (2~meV/atom), in stark contrast to the situation at 100~GPa where $Cmcm$--RbH$_5$ was at least 15~meV/atom lower in enthalpy than any other structure recovered in the EA searches.  The zero point energy (ZPE) corrections favor the mechanically stable $Cmmm$ structure by $\sim$5~meV/atom over the $Cmcm$ at 200~GPa, causing the ZPE--corrected enthalpies to fall within 5 meV/atom. The closeness of the enthalpies under pressure suggests the potential melting of the hydrogen sublattice, and quantum fluid--like behavior \cite{Bonev2004}.\\[2ex]

\begin{figure} [h!]
\begin{center}
\includegraphics[width=\columnwidth]{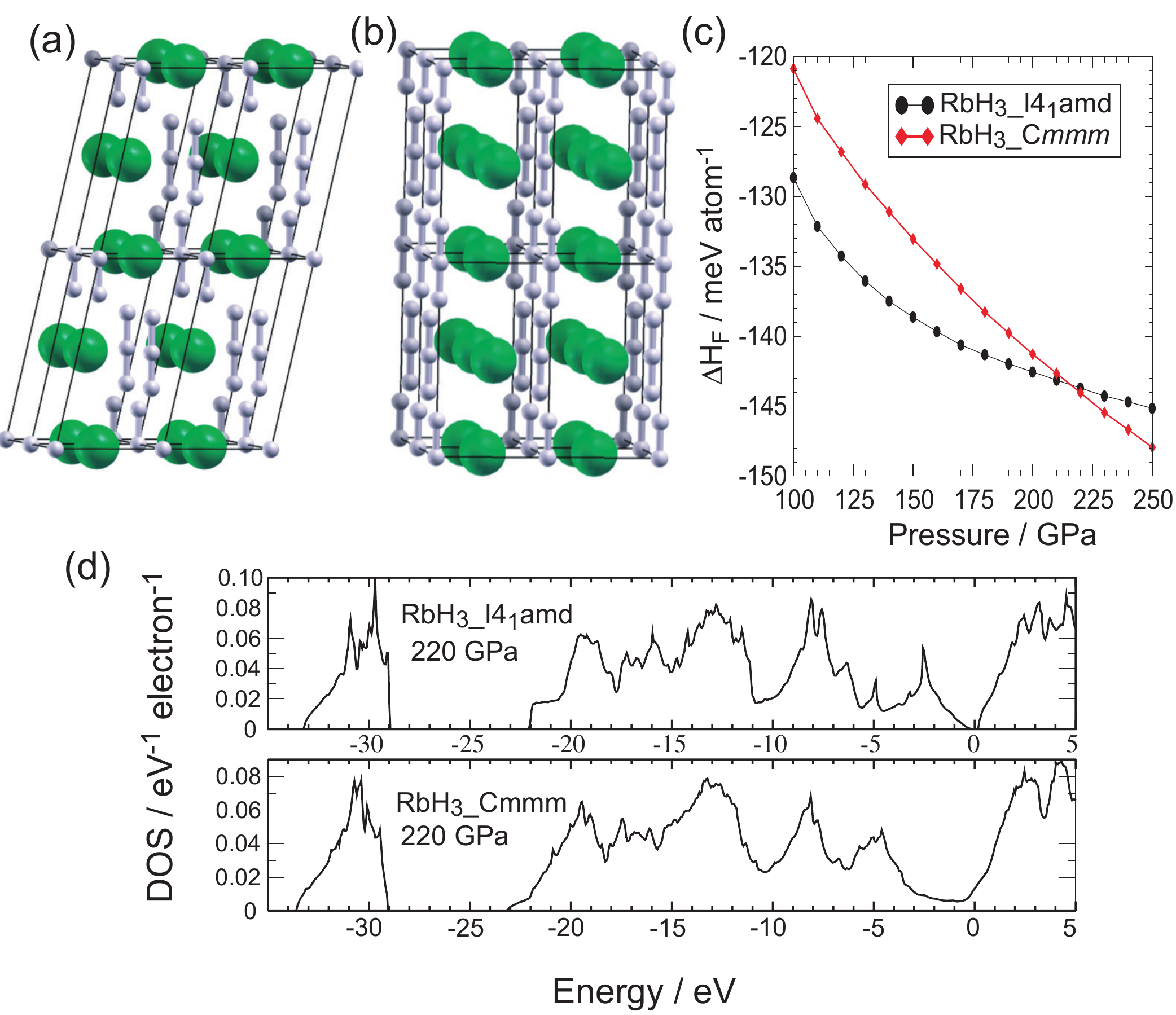}
\end{center}
\caption{Supercells of the lowest enthalpy RbH$_{3}$ phases recovered at (a) 150~GPa ($I4_1/amd$--symmetry), and (b) 250~GPa ($Cmmm$--symmetry); and (c) their enthalpies of formation at various pressures. (d) Total densities of states of (top) $I4_1/amd$--RbH$_{3}$ at 220~GPa, and (bottom) $Cmmm$--$RbH_{3}$ at 220~GPa. The Fermi levels are set to zero.   
\label{fig:RbH3_Figure}}
\end{figure}

\noindent\textbf{RbH$_3$ and RbH$_6$:  H$_3^-$ molecules and their eventual polymerization} \\
RbH$_3$, which becomes stable with respect to decomposition at 100~GPa, does not exhibit any imaginary phonon modes at 150~GPa \footnote{At 150~GPa: $a$=2.83~\AA{}, $c$=9.70~\AA{} with the atoms at the following Wyckoff sites: H $4e$ (0.500   0.250   0.217), H $2b$ (0.500   0.250   0.125), and Rb $2a$ (0.000   0.250   0.875)}. By 220~GPa, this $I4_1/amd$--symmetry structure (shown in Fig.\ \ref{fig:RbH3_Figure}(a)) overtakes RbH$_5$ to become the most stable polyhydride. Its reign, however, is shortlived as the (mechanically stable) $Cmmm$--RbH$_3$ structure in Fig.\ \ref{fig:RbH3_Figure}(b) becomes enthalpically preferred at 220~GPa \footnote{At 220~GPa: $a$=4.65~\AA{}, $b$=2.71~\AA{}, $c$=2.70~\AA{} with the atoms at the following Wyckoff sites: H $2g$ (0.188   0.000  0.000), H $1a$ (0.000   0.000   0.000), and Rb $1c$ (0.500   0.000   0.500)}, see Fig.\ \ref{fig:RbH3_Figure}(c). The difference between the ZPE corrections for the two phases was negligible at 220~GPa (within 1 meV/atom).

Both RbH$_3$ configurations are composed of Rb$^+$ cations and H$_3^-$ anions spread about a face centered cubic lattice, which is distorted somewhat in the $I4_1/amd$ structure.  The shortest intra-- and inter--H$_3^-$ H--H distances are quite similar in the two, measuring 0.87 and 1.47~\AA{}, respectively, at 220~GPa. The biggest qualitative difference between them is the stacking within the lattice along the axis of the H$_3^-$ units: an alternating/repeating sequence of Rb$^+$ and H$_3^-$ is found in the $I4_1/amd$/$Cmmm$ arrangements. The emergent layering of Rb$^+$ and H$_{3}^{-}$ in the latter is shown in Fig.\ \ref{fig:RbH3_Figure}(b), wherein the layers lie in a plane perpendicular to the page. The $Cmmm$ structure is metallic at 220~GPa, whereas the $I4_1/amd$ is a semi--metal as shown in Fig.\ \ref{fig:RbH3_Figure}(d). 
 
\begin{figure} [h!]
\begin{center}
\includegraphics[width=\columnwidth]{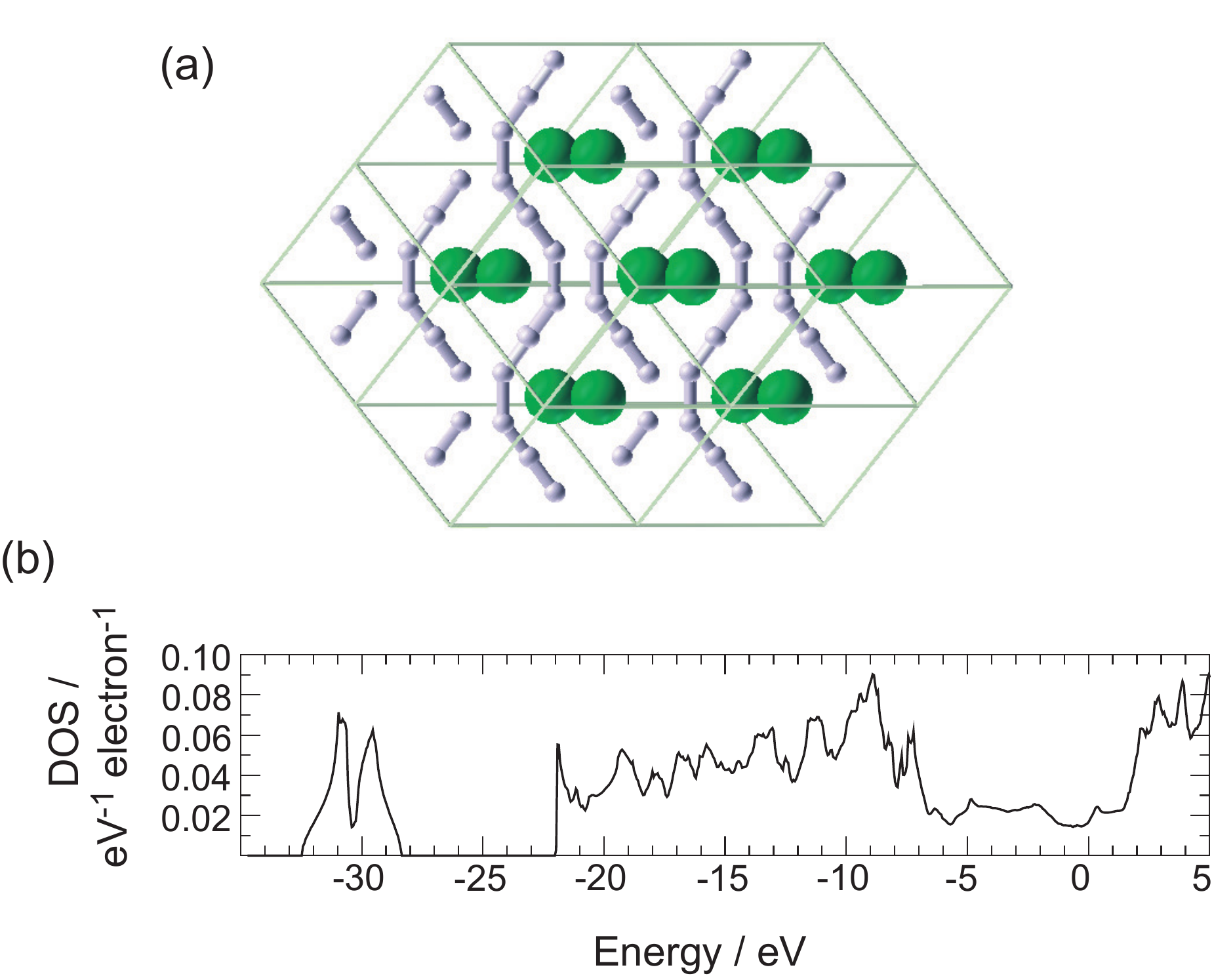}
\end{center}
\caption{(a) Supercell, and (b) total density of states  of the lowest enthalpy I$mma$--RbH$_{6}$ phase recovered at 250~GPa. The Fermi level is set to zero.   
\label{fig:RbH6_Figure}}
\end{figure}

At 250~GPa, both RbH$_3$ and RbH$_9$ are still stable but the convex hull for RbH$_n$ with $n$=3--9 flattens to the extent that I$mma$--RbH$_6$ shown in Fig.\ \ref{fig:RbH6_Figure}(a) becomes viable as well \footnote{At 250~GPa: $a$=6.54~\AA{}, $b$=3.70~\AA{}, $c$=4.75~\AA{} with the atoms at the following Wyckoff sites: H $4f$ (0.244   0.747   0.155), H $4f$ (0.099   0.391   0.485), H $4f$ (0.888   0.607   0.794), and Rb $2e$ (0.000   0.121   0.750)}. Interestingly, the hydrogen sublattice of the RbH$_6$ phase can no longer be thought of as being made up of molecular fragments, but instead of arrangements of H$_3^-$ units brought together to form what are essentially chains, ie.\ (H$_3^-$)$_\infty$. At this pressure RbH$_6$, whose DOS is illustrated in Fig.\ \ref{fig:RbH6_Figure}(b), is a good metal.  Whereas the HOMO of a symmetric H$_3^-$ is localized on the two ends of the molecule (see the SI), every hydrogen atom contributes about the same amount to the projected DOS just below the Fermi level in RbH$_6$, suggesting that a more delocalized valence electron distribution is forming along (H$_3^-$)$_\infty$.

\begin{figure} [h!]
\begin{center}
\includegraphics[width=\columnwidth]{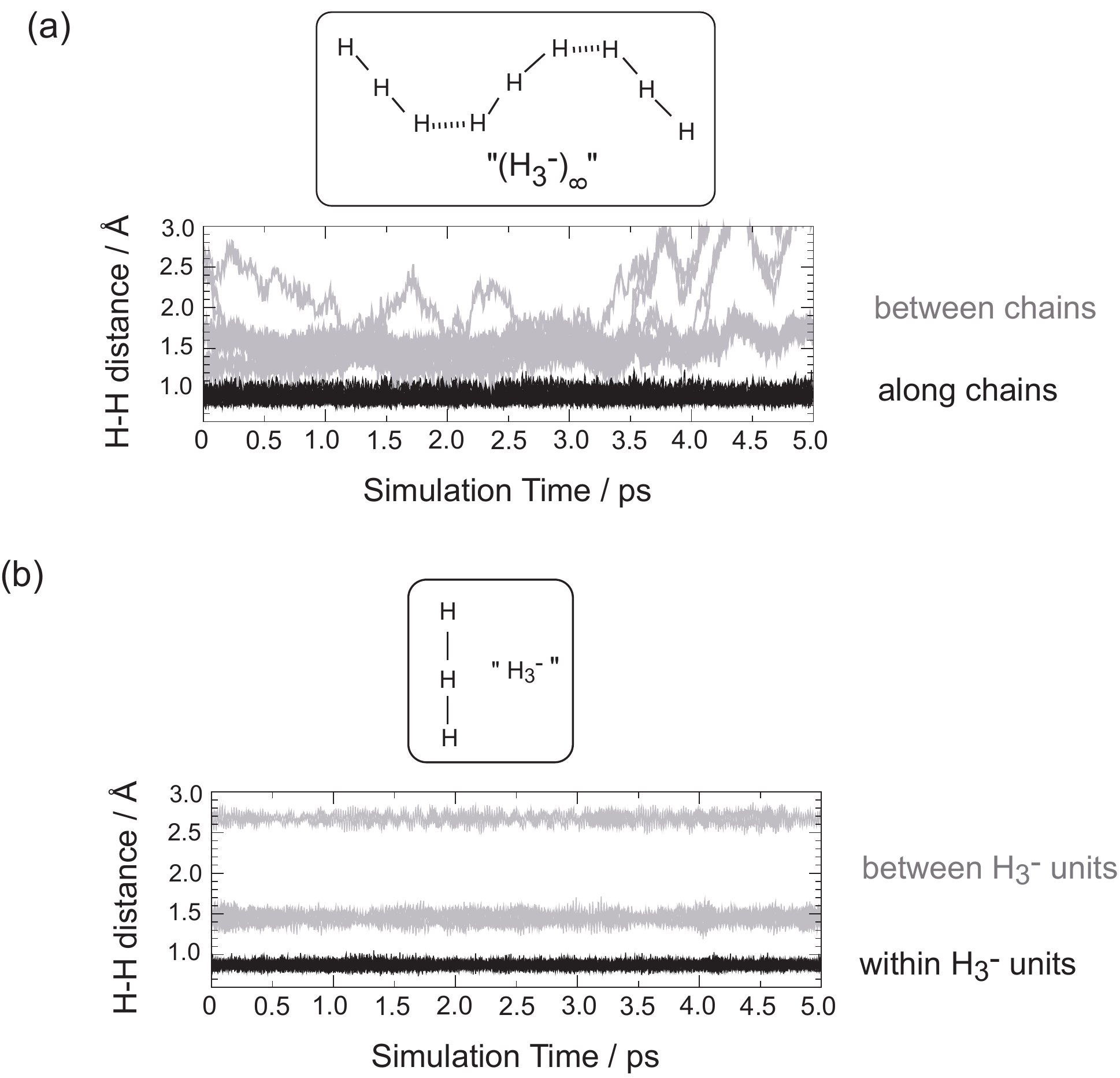}
\end{center}
\caption{The evolution of the H--H distances within the  (a) (H$_3^-$)$_\infty$ ``chains'' and the (b) H$_3^-$ molecules (black lines), and between them (gray lines) during a 5~ps molecular dynamics simulation of (a) I$mma$--RbH$_6$ and (b) $Cmmm$--RbH$_3$ at 250~GPa. 
\label{fig:RbH36_MDFigure}}
\end{figure}

Phonon calculations (in the harmonic approximation) on RbH$_6$ suggest the structure is unstable --- a couple of large--amplitude imaginary frequencies are found even at the $\Gamma$--point. Visual inspection of the soft modes indicates that they correspond to collective motions of the hydrogen atoms along the chains. To investigate this further, molecular dynamics (MD) simulations were run on a supercell of I$mma$--RbH$_6$. The evolution of a number of interatomic H--H distances from the 250~GPa--optimized RbH$_6$ structure throughout a 5~ps NVT MD simulation is shown in Fig.\ \ref{fig:RbH36_MDFigure}(a). The trajectories are shaded to illustrate whether the H--H distances corresponded to those along the constituent ``chains'' (black), or between them (gray). A similar plot is provided for RbH$_3$, where the intramolecular H--H distances are all slightly under 1.0~\AA{}, and the intermolecular ones  measure 1.5~\AA{} or higher. These distances oscillate only slightly about their equilibrium value, see Fig.\ \ref{fig:RbH36_MDFigure}(b). In RbH$_6$, on the other hand, the black and gray lines overlap at about 1.1~\AA{}, suggesting that the hydrogen sublattice is polymerizing.
\begin{figure} [h!]
\begin{center}
\includegraphics[width=\columnwidth]{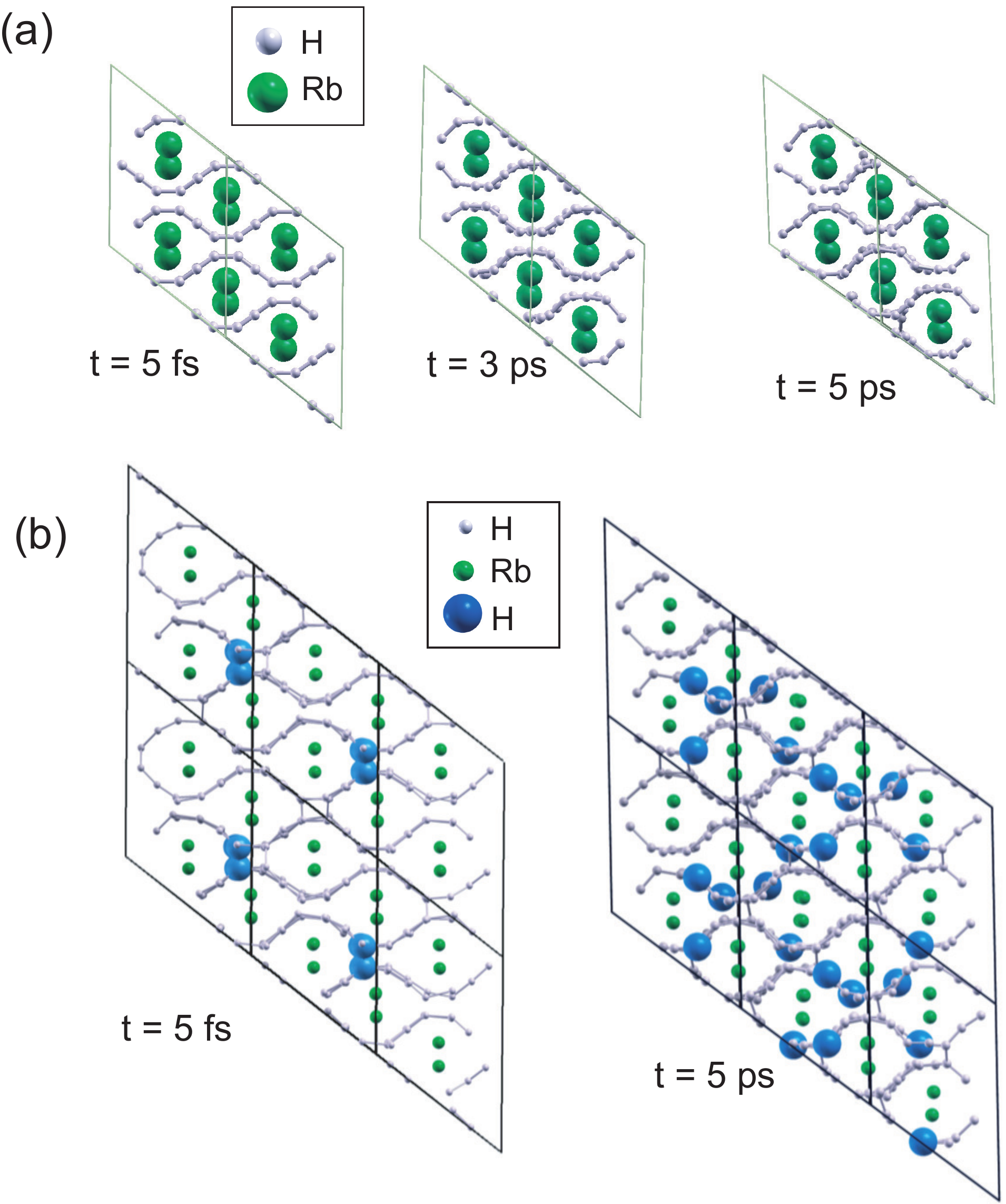}
\end{center}
\caption{(a) Optimized structures of snapshots taken from a 5~ps molecular dynamics simulation of I$mma$--RbH$_6$ at 250~GPa, and (b) the initial and final structures. In (b), six hydrogen atoms, and their images, are enlarged and highlighted in a $1 \times 2 \times 2$ represention of the simulation cell to show their displacement over the course of the simulation.
\label{fig:RbH36_MDstructures}}
\end{figure}%

Optimized snapshots from the MD simulation, wherein a ``bond'' has been drawn between two hydrogen atoms if their interatomic distance measures less than 1.2~\AA{}, are provided in Fig.\ \ref{fig:RbH36_MDstructures}(a). At first sight it may appear that each configuration maintained the distinct chain--like motifs in the original structure, but the main difference between them lies in the relative positions of the hydrogen atoms within the chains. For example, at t~=~5~fs they are eclipsed along the axis perpendicular to the page, whereas at 3 and 5~ps they are staggered. All three of the optimized structures shown in Fig.\ \ref{fig:RbH36_MDstructures}(a) are within dynamical enthalpies (3~meV/atom) of each other.   

The presence of multiple energetically degenerate local minima  and the thermally accessible barriers between them (as indicated by the short molecular dynamics run) suggest the hydrogen sublattice may even exhibit liquid--like behavior \cite{Bonev2004}, akin to that predicted for the high--pressure forms of PbH$_4$ \cite{Zaleski-Ejgierd:2011a}. This is further corroborated by noting the rise in inter--chain H--H distances toward the end of the MD--simulation in Fig.\ \ref{fig:RbH36_MDFigure}(a) is caused not by chains moving far apart from each other, but by the increase in interatomic distance between two atoms on adjacent chains as they flow. In Fig.\ \ref{fig:RbH36_MDstructures}(b), the initial and final positions of select hydrogen atoms are enlarged to show their movement during the simulation. On average, each hydrogen atom moved 1.9~\AA{} (with a standard deviation, $\sigma$, of 1.2) while each Rb atom moved only 0.07~\AA{} ($\sigma$=0.02). Furthermore, the distances between and within the H$_3^-$ building blocks of the (H$_3^-$)$_\infty$ chains  (the solid vs.\ dashed lines in Fig.\ \ref{fig:RbH36_MDFigure}(a)) are indistinguishable from one another. Since these distances measure 0.99~\AA{} and 0.87~\AA{} in the original I$mma$--RbH$_6$ structure, only a 12\% difference, this suggests that the thermal vibrations of the H atoms about the chain are enough to circumvent the distinction of the two characteristic H--H distances. The hydrogenic sublattice in this structure can be thought to consist of flowing, partially negatively charged, polymeric chains.\\[2ex]

\section{Conclusions}

Our first--principles computations predict a number of rubidium polyhydrides, RbH$_n$ $n>1$ , which become stable at pressures  as low as 2~GPa. There is a wide variety in their structures, and their hydrogenic sublattices can be viewed as being composed of H$^-$, H$_2$ and H$_3^-$ units. We note the similarities with polyiodide compounds whose complex, and often times polymeric--like lattices can be understood in terms of interactions between the I$^-$, I$_2$, as well as the symmetric and assymmetric linear (or nearly) I$_3^-$ building blocks \cite{Svensson:2003a}.

RbH$_5$, the most stable structure between 15-220~GPa contains Rb$^+$ and H$_2$ molecules, along with the symmetric \hm\ motifs --- the simplest example of a 3c-4e bond. The broadened Rb $4p$ bands hybridize with hydrogenic bands, and the overlap of the Rb $4s/4p$ orbitals makes this phase an insulator. Compress further and a metallic RbH$_3$ phase made up Rb$^+$ and H$_3^-$ building blocks becomes the most stable polyhydride. At 250~GPa, alternative forms of the hydrogen sublattice emerge in RbH$_6$ in the form of polymeric chains. Molecular dynamics simulations suggest the hydrogen sublattice in RbH$_6$ may exhibit one dimensional liquid--like behavior. 

\section{Computational Details} \label{sec:comp}

The structural searches were performed using the open--source EA XtalOpt Release 7 \cite{Zurek:2011f} (http://xtalopt.openmolecules.net/), and the parameter set suggested in Ref.\ \cite{Zurek:2011a}. EA searches were run on simulation cells with two RbH$_n$ formula units at 2, 10, 30, 100, 150 and 250~GPa.  The lowest enthalpy structures from each search were relaxed in the pressure range from 0-250~GPa. 

First principles DFT calculations were performed with the Vienna \textit{ab--initio} Simulation Package (VASP) version 4.6.31 \cite{Kresse:1993a}. The projector augmented wave (PAW) method \cite{Blochl:1994a} was used to treat the core states, and a plane--wave basis set with an energy cutoff of 600~eV was employed. An energy cutoff of 500~eV was used in the structural searches. The Rb $4s/4p/5s$ electrons were treated explicitly in all of the calculations. The gradient--corrected exchange and correlation functional of Perdew--Burke--Ernzerhof (PBE) \cite{Perdew:1996a} was adopted, and the $k$--point grids were generated using the $\Gamma$-centered Monkhorst--Pack scheme.  The number of divisions along each reciprocal lattice vector was chosen such that the product of this number with the real lattice constant was 20~\AA{} in the structural searches and 40~\AA{} otherwise.

Phonons and thermodynamic properties of select RbH$_3$, RbH$_5$, RbH$_6$, and RbH$_9$ phases were calculated using the PHON \cite{phon} package. The supercells used for the phonon calculations were chosen such that the number atoms in the simulation cell was always between 96 and 216 atoms. We have neglected the zero point energy (ZPE) corrections to the enthalpy in most of our discussion. Since the importance of ZPE contributions when taking into account the stability of systems with light atoms has been well--documented in the literature \cite{Pickard:2007a}, we provide two examples in the text of how the inclusion of the ZPE term influences the stability RbH$_3$ and RbH$_5$ at 200--220~GPa. Overall, our assessment is that at the pressures we focus on, the differences in ZPE between RbH$_n$ structures are never large.

In order to assess the mechanical stability of RbH$_6$, ab--initio molecular dynamics (MD) simulations were run on RbH$_6$ using VASP. The MD simulations were performed under constant volume and temperature conditions (NVT) for 5 ps with 0.5 fs timesteps. The initial velocities were initialized randomly according to a Maxwell--Boltzmann distribution such that the starting temperature of the system was 400~K and the temperature was maintained through intermittent velocity rescaling. A 96--atom simulation cell was used for RbH$_6$ and a 128--atom simulation cell for the complementary RbH$_3$ simulation.

The band structures of select phases were calculated using the tight--binding linear muffin--tin orbital (TB--LMTO) method \cite{Andersen:1975,Andersen:1984}, and for hp--RbH$_5$ $N$MTO methods (MTOs of order $N$) \cite{Zurek:2005a, Andersen:2000}. In the TB--LMTO calculations the VWN \cite{Vosko:1980} local exchange correlation potential was used along with the Perdew-Wang \cite{Perdew:1986} generalized gradient approximation. Scalar relativistic effects were included. The structural parameters of hp--RbH$_5$ were taken from those optimized with VASP at 100~GPa, and four different types of empty interstitial spheres were inserted, each with a single $s$ function. The calculations utilized 2768 irreducible points in the tetrahedron $k-$space integrations \cite{Blochl:1994}. The TB--LMTO densities of states showed good agreement to those obtained with VASP (see the SI). The fat bands, whose width is proportional to a particular orbital character, are also provided in the SI.

The band structure with a full $N$MTO basis (Rb $4s$, $4p$, $5d$ and H $1s$, $2p$) of RbH$_5$ at 100~GPa corresponded well with the bands of Fig.\ \ref{fig:hooper-fig4}(a). The ``Rb 4p + H$_2$ $\sigma$ + H$_3^-$ bonding'' bands in this figure were computed with an $N$MTO minimal basis consisting of $s$ and $p$ functions on all Rb atoms, an $s$ function on every second H$_2$ atom, and an $s$ function on the central H atom in an \hm\ unit. The ``H$_3^-$ non--bonding'' bands were computed with an $s$ function on one of the terminal hydrogens in each \hm\ unit. In the SI, we show these bands along with the energy meshes used in the calculations.

\section*{Acknowledgements}
We acknowledge the NSF (DMR-1005413) for financial support, and the Center for Computational Research at SUNY Buffalo for computational support.

%\bibliography{pressure,vasp,hydrides,zurek,hions,lmto,nmto} %multicenter,RbH5_otherrefs,Pio}% Produces the bibliography via BibTeX.

\providecommand*{\mcitethebibliography}{\thebibliography}
\csname @ifundefined\endcsname{endmcitethebibliography}
{\let\endmcitethebibliography\endthebibliography}{}

\clearpage

\noindent \textbf{Table of Contents Text:} \\ 
\textbf{A pressing motif:} At 15~GPa RbH$_5$ is predicted to become stable with respect to decomposition into H$_2$ and RbH. The unit cell contains Rb$^+$ cations, H$_2$ molecules, and the symmetric linear \hm\ unit --- the simplest example of a three--center four--electron bond. At higher pressures \hm\ is further stabilized in an RbH$_3$ phase which metallizes at 220~GPa. 

\noindent\textbf{Keywords:} hydrides, alkali metals, ab--initio calculations, high--pressure chemistry, bond theory

\begin{figure} [h!]
\begin{center}
\includegraphics[width=0.8\columnwidth]{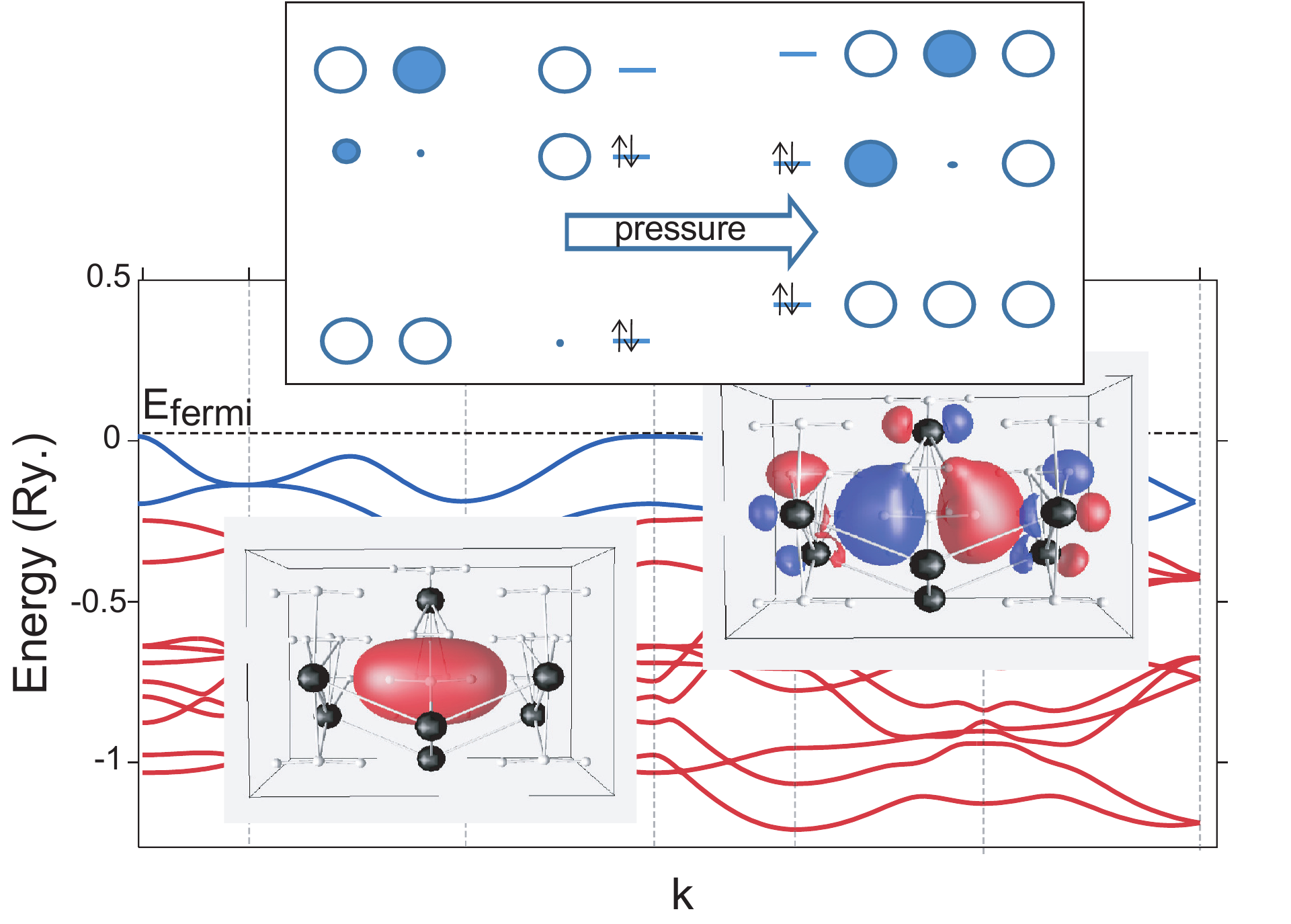}
\end{center}
\end{figure}

\end{document}